\documentclass[%
 aip,
 amsmath,amssymb,
 reprint,%
]{revtex4-1}
\usepackage{color}
\usepackage{graphicx}
\usepackage{dcolumn}
\usepackage{bm}

\usepackage[utf8]{inputenc}
\usepackage[T1]{fontenc}
\usepackage{mathptmx}
\usepackage{etoolbox}
\usepackage{url}
\usepackage{hyperref}
\hypersetup{
    colorlinks=true,
    linkcolor=blue,
    urlcolor=blue,
    citecolor = black,
    pdftitle={Overleaf Example},
    pdfpagemode=FullScreen,
    }

\makeatletter
\def\@email#1#2{%
 \endgroup
 \patchcmd{\titleblock@produce}
  {\frontmatter@RRAPformat}
  {\frontmatter@RRAPformat{\produce@RRAP{*#1\href{mailto:#2}{#2}}}\frontmatter@RRAPformat}
  {}{}
}%
\makeatother
\begin{document}

\preprint{AIP/123-QED}

\title[The following article has been submitted to Chaos. After it is published, it will be found at \href{https://aip.scitation.org/journal/cha}{Chaos}.]{Perspectives on the importance of complex systems in understanding our climate and climate change - The Nobel Prize in Physics 2021}
\author{Shraddha Gupta}
\email{shraddha.gupta@pik-potsdam.de}
\affiliation{Potsdam Institute for Climate Impact Research (PIK) -- Member of the Leibniz
Association, Telegrafenberg A56, Potsdam, 14473, Germany.}
\affiliation{Department of Physics, Humboldt University at Berlin, Newtonstraße 15, Berlin, 12489, Germany. }
\author{Nikolaos Mastrantonas}%
 \email{nikolaos.mastrantonas@ecmwf.int}
\affiliation{ 
European Centre for Medium-Range Weather Forecasts, Shinfield Park,
Reading, RG2 9AX, United Kingdom.
}%
\affiliation{Interdisciplinary Environmental Research Centre, Technische Universität Bergakademie Freiberg (TUBAF), Brennhausgaße 14, 09599 Freiberg, Germany.}

\author{Cristina Masoller}
 \homepage{http://www.fisica.edu.uy/~cris/}
\affiliation{Departament de Física, Universitat Politecnica de Catalunya, Sant Nebridi 22, 08222 Terrassa, Spain.
}%

\author{Jürgen Kurths}
 \homepage{https://www.pik-potsdam.de/members/kurths/homepage}
\affiliation{Potsdam Institute for Climate Impact Research (PIK) -- Member of the Leibniz
Association, Telegrafenberg A56, Potsdam, 14473, Germany.}
\affiliation{Department of Physics, Humboldt University at Berlin, Newtonstraße 15, Berlin, 12489, Germany. }
\affiliation{Institute of Information Technology, Mathematics and Mechanics, Lobachevsky University of Nizhny Novgorod, Nizhnij Novgorod 603950, Russia.}

\date{\today}

\begin{abstract}

The Nobel Prize in Physics 2021 was awarded to Syukuro Manabe, Klaus Hasselmann and Giorgio Parisi for their "groundbreaking contributions to our understanding of complex systems” including major advances in the understanding of our climate and climate change. In this perspective article, we review their key contributions and discuss their relevance in relation to the present understanding of our climate. We conclude by outlining some promising research directions and open questions in climate science.

\end{abstract}

\maketitle

\begin{quotation}

Classic complex systems, as coupled pendula, nonlinear circuits or lasers, are typically constituted by a few elements or subsystems, whose dynamical behaviour and interactions are nonlinear and may involve memory effects. Due to these properties, they are able to generate rich and even chaotic dynamics, i.e. long-term predictions fail.  In contrast, “complicated” systems can be very large, but their governing equations are linear. Thus, their system’s behavior can be understood by using a “reductionist” approach, and it can be well predicted from the behavior of the individual subsystems. However, many real systems are complex and they consist of many components, as power grids or the human brain. Our climate and Earth system as a whole is another outstanding example of such a large complex system. Additionally, it covers a broad range of scales in space and time.  Hence, it cannot be appropriately described and understood by using the reductionist approach but requires advanced techniques from complex systems science. Here, we discuss how the pioneering works of Syukuro Manabe, Klaus Hasselmann and Giorgio Parisi (Physics Nobel Prize 2021) have given us crucial insights for understanding the Earth climate, and basic underlying mechanisms of climate change and what are recent directions in this very active field of research.
\end{quotation}


\section{\label{sec:Intro}Introduction}

\begin{figure}
    \centering
    \includegraphics[scale=0.8]{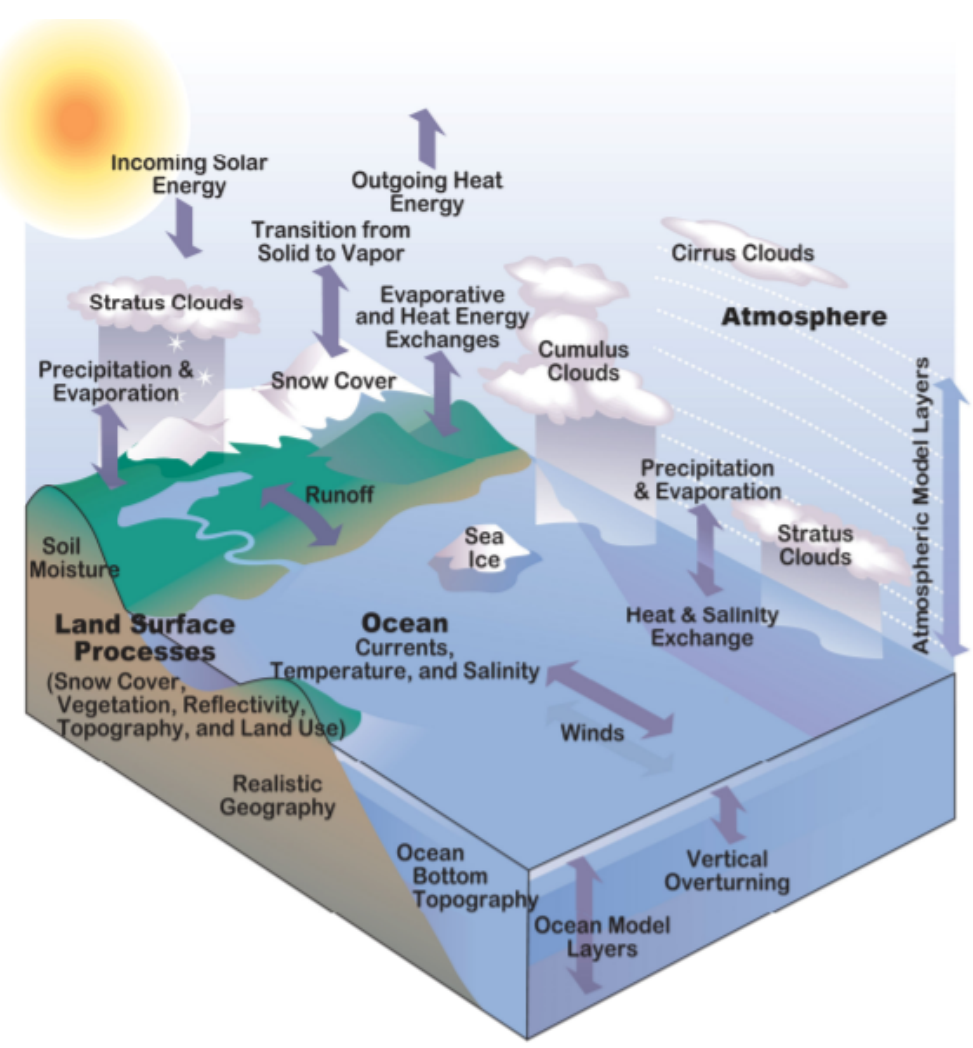}
    \caption{\small Schematic representation of the interaction between the different components of the Earth system from the Community Climate System Model (CCSM), developed by the University Corporation for Atmospheric Research (UCAR)\cite{UCAREORa4:online}. \copyright \,2022 UCAR. Illustration by Paul Grabhorn.}
    \label{fig:my_label}
\end{figure}

The Earth system is a highly complex system whose components (e.g., atmosphere, land, cryosphere) and interactions, schematically represented in Fig. \ref{fig:my_label}, are driven by various physical, biological, and chemical processes, which act with different spatial and temporal scales.

To be able to predict the Earth’s weather (in the next few days) and climate (in the next months and longer), the international community has dedicated a great number of resources to develop advanced Earth System Models (e.g.\cite{UCAREORa4:online,ECMWF}). These models numerically simulate the Earth’s climate by incorporating in best detail all the known processes that take place from the depth of the oceans to the highest levels of the Earth’s atmosphere. The impressive knowledge of the Earth’s representation and the understanding of climate change has been reached after a long history of model development, with milestones that substantially improved our understanding.  

Already by the end of the 19th century, Svante Arrhenius (Chemistry Nobel Prize 1903) understood the role of greenhouse gases in the atmosphere in maintaining the Earth’s temperature at the observed value, by absorbing the long-wave radiation emitted from the Earth’s surface\cite{Arrhenius1896}. He was the first to quantitatively associate rising temperatures with the increase of greenhouse gases.  

One of the first to conceive the notion that the laws of physics could be used for weather prediction was the Norwegian meteorologist Vilhelm Bjerknes, who considered weather forecasting as an initial value problem of mathematical physics\cite{Bjerkenes1904}. Inspired by Bjerknes, in the 1920s, Lewis Fry Richardson confronted the impossible calculus of differential equations by replacing them with more tractable  "finite difference" equations\cite{LFRichardson}. His method of computation, although tedious at that time, laid the foundations of present-day numerical weather prediction models with the advent of electronic computers. In 1950, John von Neumann was working together with
Jule Gregory Charney, and produced the first numerical weather forecast\cite{Neumann1950} using the first programmable computer, the ENIAC. 

In 1963, Edward Lorenz\cite{Lorenz63} discovered deterministic chaos in a model he developed for studying convective process in the atmosphere. His equations were a strongly simplified version of the ones derived by Barry Saltzman\cite{Saltzman1962} to study the Rayleigh-B\'{e}nard convection. His findings created a new paradigm in science by imposing a limit on the predictability of the weather, as extremely small errors in the initial state amplify rapidly and lead to large uncertainties in the forecasts, when they are longer than about 10 days. In the following decades, unprecedented advances in computing power and observations from different sources (satellites, radar systems, buoys, etc.) and the subtle use of nonlinear data assimilation as well as new numerical techniques have enabled the scientific community to build highly sophisticated weather models called the General Circulation Models (GCMs)\cite{WhatisaG29:online,ClimateM19:online}. 
 
Humans have strongly influenced our Earth System due to their various activities. Although in 1861 John Tyndall\cite{Tyndall1861} had already identified CO$_2$ as a greenhouse gas, it was only in the late 20$^\textrm{th}$ century that human influence on climate variability was quantitatively measured. The weight of evidence on detectable anthropogenic influence on the course of climate have accumulated rapidly over the past decades. The very recent assessment report of the Intergovernmental Panel on Climate Change (IPCC AR6), entitled “Climate Change 2021: the Physical Science Basis”\cite{IPCC:21} compiles the recent studies about climate change, and provides overwhelming evidence of the increase in the Earth’s surface temperature due to anthropogenic interventions, and the subsequent alterations that are expected in the occurrence of extreme weather across the globe.

In 2021 the Physics Nobel Prize recognized the importance of understanding variability which leads to various emergent phenomena in different complex systems, such as the Earth's climate. Half of the Nobel Prize was awarded to Syukuro Manabe and Klaus Hasselmann for their contributions to the understanding and modelling of the Earth's climate. The other half was awarded to Giorgio Parisi for his contributions to the theory of complex systems, which have applications not only in different areas of physics including climate science, but also in mathematics, neuroscience, machine learning and various other disciplines. 

Syukuro Manabe believed that equal emphasis should be placed on both understanding as well as predicting climate change and tried to find a relation between the increase in global temperature with that of the level of CO$_2$ in the atmosphere\cite{Manabe1967}. Klaus Hasselmann demonstrated how climate variability can be driven by stochastic short time scale weather fluctuations\cite{Hasselmann1976,Hasselmann1977}. He also proposed a systematic framework to compare climate models and observations, to distinguish, in climate signals, the imprints of natural variability from those caused by the human-induced increase of greenhouse gases\cite{Hasselmann1993,Hasselmann1997}.  

In the following sections we first summarize some main contributions of the three Nobel awardees to climate science, and how they built upon previous findings. Then, we discuss how their works are still advancing our knowledge on the Earth System and, finally, we conclude with some open questions regarding the future climate of our planet, promising directions as well as the role the scientific community should take to shape its future.

\section{{{Manabe and Hasselmans' Contributions to Climate Modelling
}}}\label{sec:NWP}

The importance of the atmosphere, in particular the greenhouse gases, in making our planet habitable by maintaining a livable temperature is well known today. 

From the very first zero-dimensional model introduced by Arrhenius, to the currently used state-of-the-art GCMs, there is a long history of contributions that can be seen from the perspective of model hierarchies \cite{ModelHierarchies}. The initial developments made use of conceptual energy-balance models. Such models improved the understanding of the complexity of the Earth System, and this increased understanding  motivated further improvement of modelling the Earth climate, with the addition of more processes and spatiotemporal scales and interactions. In this context, the works of Syukuro Manabe and Klaus Hasselmann are of fundamental importance.

\subsection{Manabe's 1D Radiative-Convective Model}

\begin{figure}
    \centering
    \includegraphics[scale=0.63]{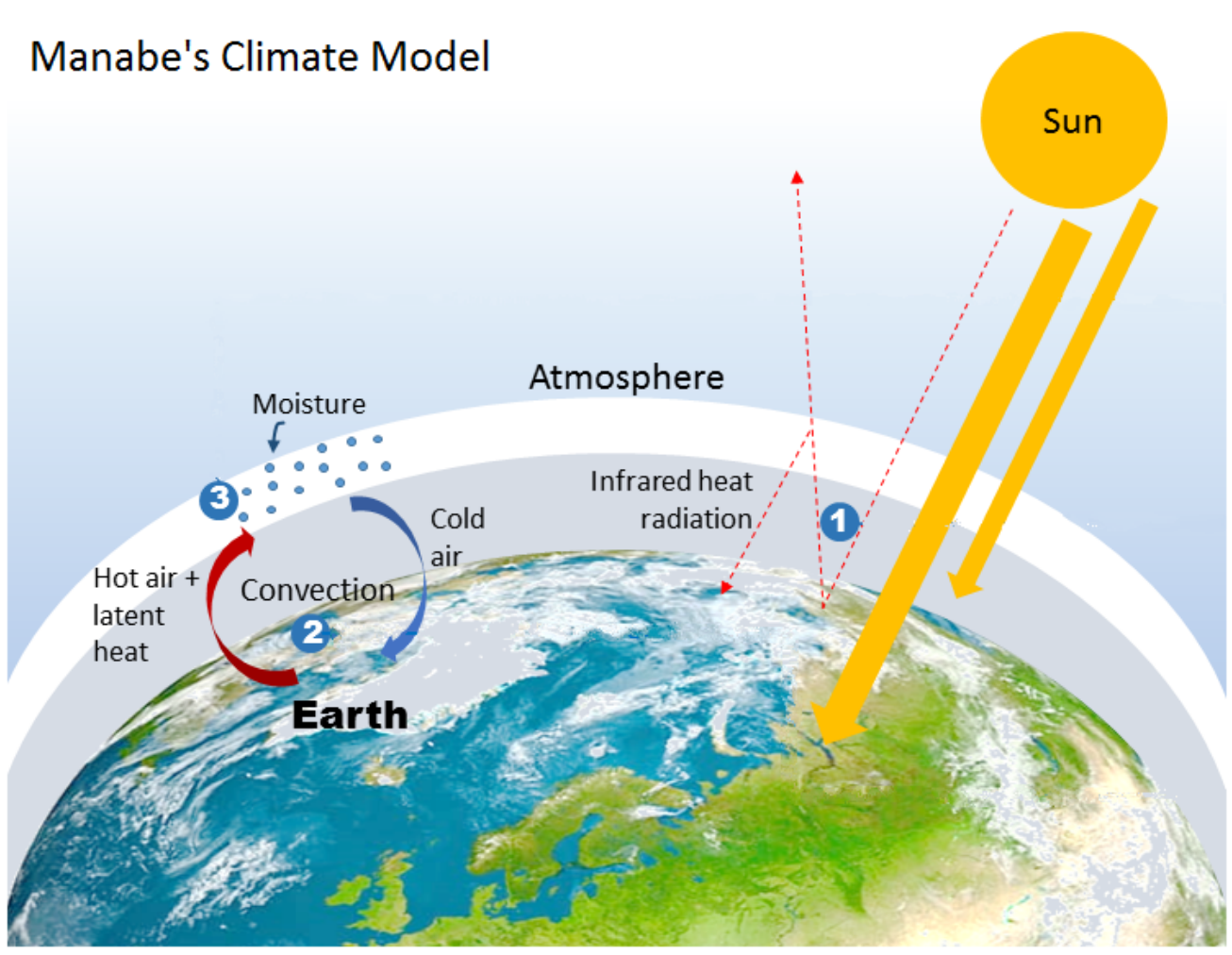}
    \caption{\small Schematic representation of Syukuro Manabe's climate model, which was the first to incorporate the interaction between radiation balance and the vertical transport of air mass due to convection, in addition to the heat contributed by the water cycle. Labels (1), (2) and (3) represent the following processes: (1) Infrared heat radiation from the ground is partially absorbed in the atmosphere, warming the air and the ground. Some radiates out into space. (2) Hot air being lighter than cold air rises through convection, carrying water vapour, a powerful greenhouse gas, along with it. The concentration of water vapour increases as the air becomes warmer. (3) Higher up, where the atmosphere is colder, cloud drops form and release the latent heat stored in the water vapour.}
    \label{fig:manabemodel}
\end{figure}

\begin{figure}[ht]
    \centering
    \includegraphics[scale=0.8]{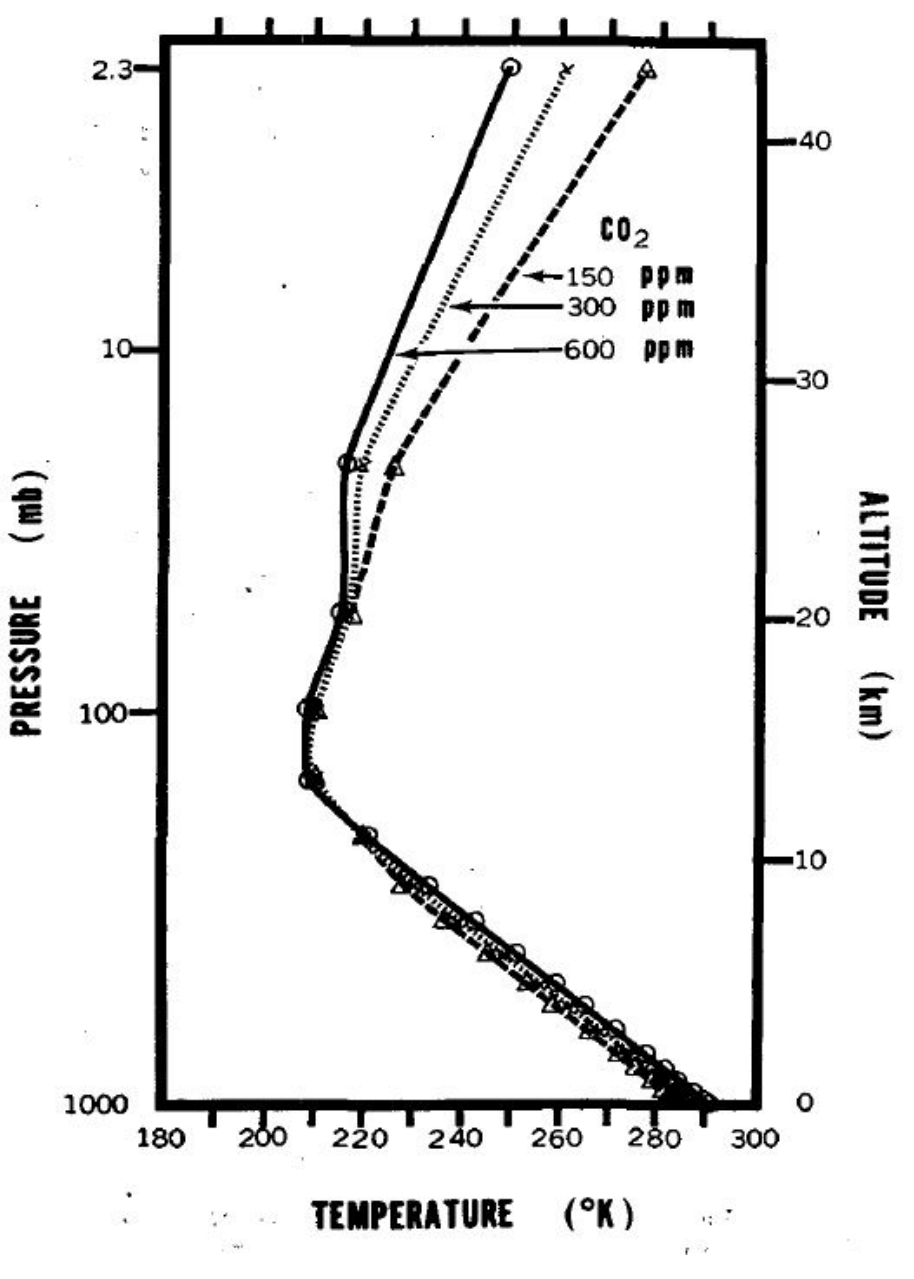}
    \caption{\small Vertical profiles of temperature in radiative-convective equilibrium is shown for three different atmospheric concentrations of carbon dioxide, that is, 150 ppm (part per million), 300 ppm, and 600 ppm by volume as dashed, dotted and solid lines, respectively.  From \citeauthor{Manabe1967}\cite{Manabe1967}. Published 1967 by the American Meteorological Society.}
    \label{fig:ManabeCO}
\end{figure}

Syukuro Manabe realized that the heating of the Earth’s surface would warm the air in close contact with the Earth’s surface, generating convection. As convection proceeds, warm air (water vapour) rises and adiabatically cools, leading to condensation of water vapour and release of its latent heat. Taking this convective adjustment into consideration, in 1967 \citet{Manabe1967} modelled the atmosphere as a one-dimensional vertical column with an initial profile of relative humidity and greenhouse gases. This profile evolved with time according to the dynamics of radiative transfer and upward convection of water vapour (Fig. \ref{fig:manabemodel}). They found that although a change in oxygen and nitrogen levels has a negligible impact on temperature, doubling the CO$_2$ concentration leads to an increase in the global surface temperature by 2.36°C, while the temperature in the stratosphere substantially decreases (Fig. \ref{fig:ManabeCO}). This difference between the response of the troposphere and that of the  stratosphere is due to the absence of convective heating in the latter. The net cooling due to the emission and absorption of long wave radiation, mainly by CO$_2$, and the heating due to the absorption of solar radiation by ozone are the two major processes which maintain the heat balance in the stratosphere. An increase in the CO$_2$ concentration in the atmosphere would enhance the net long wave cooling thereby lowering the equilibrium temperature of the stratosphere\cite{Manabe2019}. 

It is interesting to note that in 1938 a British engineer, Guy Callendar\cite{Callendar1938}, discovered the potential impact of anthropogenic CO$_2$ emission into the atmosphere on the global climate through a simple radiative energy balance model of the Earth’s surface. Although Fritz Möller, one of Manabe’s collaborators noted serious flaws in Callendar’s approach, Manabe realised that his radiative-convective model was an excellent conceptual tool to study the greenhouse effect. Manabe's findings confirmed that the differential heating and cooling of the lower and upper atmosphere, respectively, is caused due to increased levels of the greenhouse gases. If the heating was caused by the increase in solar radiation, the entire atmosphere should have warmed up. 
   
Manabe's work was a milestone in the understanding of the complex interactions in the Earth system and the effect of CO$_2$. Not only was it was the first published paper to use a computational model and calculate the global warming and stratospheric cooling from increased CO$_2$, but it was also the first to assess the magnitude of water vapour feedback. At the same time, this work introduced two important ideas. Firstly, it considered the heat balance of the atmosphere as a whole, rather than the heat balance of the Earth’s surface\cite{Moller1963} which was a crucial advancement, as it links the different atmospheric levels with the Earth's surface thereby accounting for their interactions while computing the changes in their temperatures.  Secondly, it assumed that the changes in the atmospheric temperature will not affect the relative humidity instead of the absolute humidity as used in the previous work\cite{Manabe1964}. Thereafter, this assumption was used to model the positive temperature-water vapour feedback. The work comprehensively demonstrated that the combination of radiation and convection (radiative-convective scheme), the use of relative humidity and the incorporation of water vapour feedback could substantially improve the representation of the system, thus supporting more realistic GCMs. It is fascinating that Manabe’s rather simple conceptual 1-D model was able to quantify climate sensitivity more accurately than the more complex GCMs that were used at that time. In fact, the CO$_2$-related findings of this work, are in very close agreement with the current state-of-the-art climate models, despite the many simplifications of that study. Simplifications that were nevertheless necessary given the available computational resources.

\subsection{Manabe's 3D General Circulation Model}

Proceeding one step further, Manabe incorporated the effects of hydrology at Earth’s surface in a three-dimensional (3D) GCM with nine vertical layers resolving the atmosphere from the surface boundary layer till the stratosphere\cite{Manabe1969}. As he concluded “{\it {the interaction between the hydrology of the Earth's surface and the general circulation of the model atmosphere results in a highly realistic distribution of precipitation, evaporation, and sensible heat flux and net radiative flux at the earth's surface.}}” He then continued improving the models by increasing their resolution and the number of equations. In 1975, \citet{Manabe1975} simulated for the first time the 3D response of the hydrological cycle and temperature to increased CO$_2$ concentrations in the Earth’s atmosphere, and found that the doubling of CO$_2$ significantly intensifies the hydrological cycle of the model. All these developments finally led to the incorporation of the ocean-atmosphere interaction to the GCMs\cite{Manabe1980}.  

\subsection{Hasselmann's Model of Climate Variability}


Before Klaus Hasselmann introduced in 1976 his stochastic model\cite{Hasselmann1976} for predicting climatic variability, it was well-known that the climate system, described by its state vector ${\bf z}=(z_1,\,z_2,\,...)$ that represents climatic variables -- density, wind velocity, temperature, etc.-- at a given time at spatial grid points at different levels, could be divided into two subsystems: \textit{(1)} the rapidly-varying "weather" system ${\bf x}$ and \textit{(2)} the slowly-varying "climate" response ${\bf y}$ 
, having their own characteristic time scales $\tau_x$ and $\tau_y$, respectively.

Then, evolution of the complete climate system could be described by a set of equations:
\begin{eqnarray}
    \frac{dx_i}{dt}&=&u_i({\bf x},\,{\bf y}),\\
    \label{eq:1}
    \frac{dy_i}{dt}&=&v_i({\bf x},\,{\bf y}),
    \label{eq:2}
\end{eqnarray}
where $u_i$ and $v_i$ are nonlinear functions of the weather and climate variables. In general, the climate variables $y_i$ may be associated with sea surface temperature, ice coverage or land foliage, among others. Since the weather fluctuates on a daily basis and climate varies over times scales of months to several years or longer, $\tau_x\ll\tau_y$ and ${\bf y}$ could be set as constant in weather models. 

Although the GCMs could be used to predict the weather components, $x_i$, they could not be simulated for long enough time scales of climatic relevance due to the limitation on computational resources. As an alternative, Statistical Dynamical Models (SDMs) had been used to study climatic variability\cite{Budyko1969,Sellers1969}. However, since these models averaged out the small-time-scale weather fluctuations, they turned out to be deterministic rather than stochastic and often displayed the same asymptotic behaviour for a range of initial conditions. As a result, SDMs were not able to capture the red noise variance spectra of the observed climate data. In order to explain climatic variability in the context of SDMs, researchers incorporated external perturbations (such as changes in the solar radiation and turbidity of the atmosphere), while neglected the importance of weather fluctuations.

In his stochastic model, Hasselmann highlighted the important role of weather fluctuations in determining the climate variability. He assumed that over time scales $t\ll\tau_y$, the change in a climate variable $y_i$ from its initial state can be divided into a mean $\left<y_i\right>$ and a fluctuating part, $y_i'$. Similarly, the forcing function $v_i$ can be written as $\left<v_i\right>+v_i'$, where the rapidly changing weather components manifest as the fluctuations $v_i'$ which act as random forcing terms in the model i.e.
\begin{equation}
    \frac{dy_i'}{dt}=v_i'.
    \label{eq:3}
\end{equation}
In a seminal work in 1921, G. I. Taylor\cite{Taylor1921} showed that Eq. (\ref{eq:3}) is the continuum mechanical analogue of normal molecular diffusion or the Brownian motion.  From Eq. (\ref{eq:3}), it follows that the covariance, $\left<y_i'y_i'\right>$, grows linearly with time over climatic time scales, i.e.,
\begin{equation}\label{covariance}
    \left<y_i'y_j'\right>=2D_{ij}t
\end{equation}
where 
\begin{equation}
    D_{ij}=\frac{1}{2}\int_{-\infty}^\infty P_{ij}(\tau)d\tau
\end{equation}
is the diffusion coefficient and $P_{ij}=\left<v_i'(t+\tau)v_j'(t)\right>$ denotes the covariance function of the forcing. In terms of the spectral representation of Eq. \ref{covariance}, the diffusion coefficient can then be simply expressed in terms of spectral density of the $P_{ij}$, $F_{ij}(\omega)$ at zero frequency, by considering only the first term of the Taylor expansion of the spectrum. 
\begin{equation}
    D_{ij}=\pi F_{ij}(0)
\end{equation}
It follows that for sufficiently small frequencies $\tau_y^{-1}<\omega<\tau_x^{-1}$, $F_{ij}(\omega)$ approaches $F_{ij}(0)$, and for $F_{ij}(0)\neq0$, the spectrum of input $v_i$ is white.

Hasselmann thus argued that the climate system is analogous to the Brownian motion problem -- exhibiting the same random-walk response characteristics as large particles interacting with an ensemble of much smaller particles. As a result, it is possible to write a Fokker-Planck equation for the climate probability distribution $p({\bf y},t)$, in which the diffusion term $D_{ij}$ is governed by the fluctuating weather components and the direct internal coupling is included through the propagation term $\hat{v_i}$.

\begin{figure}
    \centering
    \includegraphics[scale=0.5]{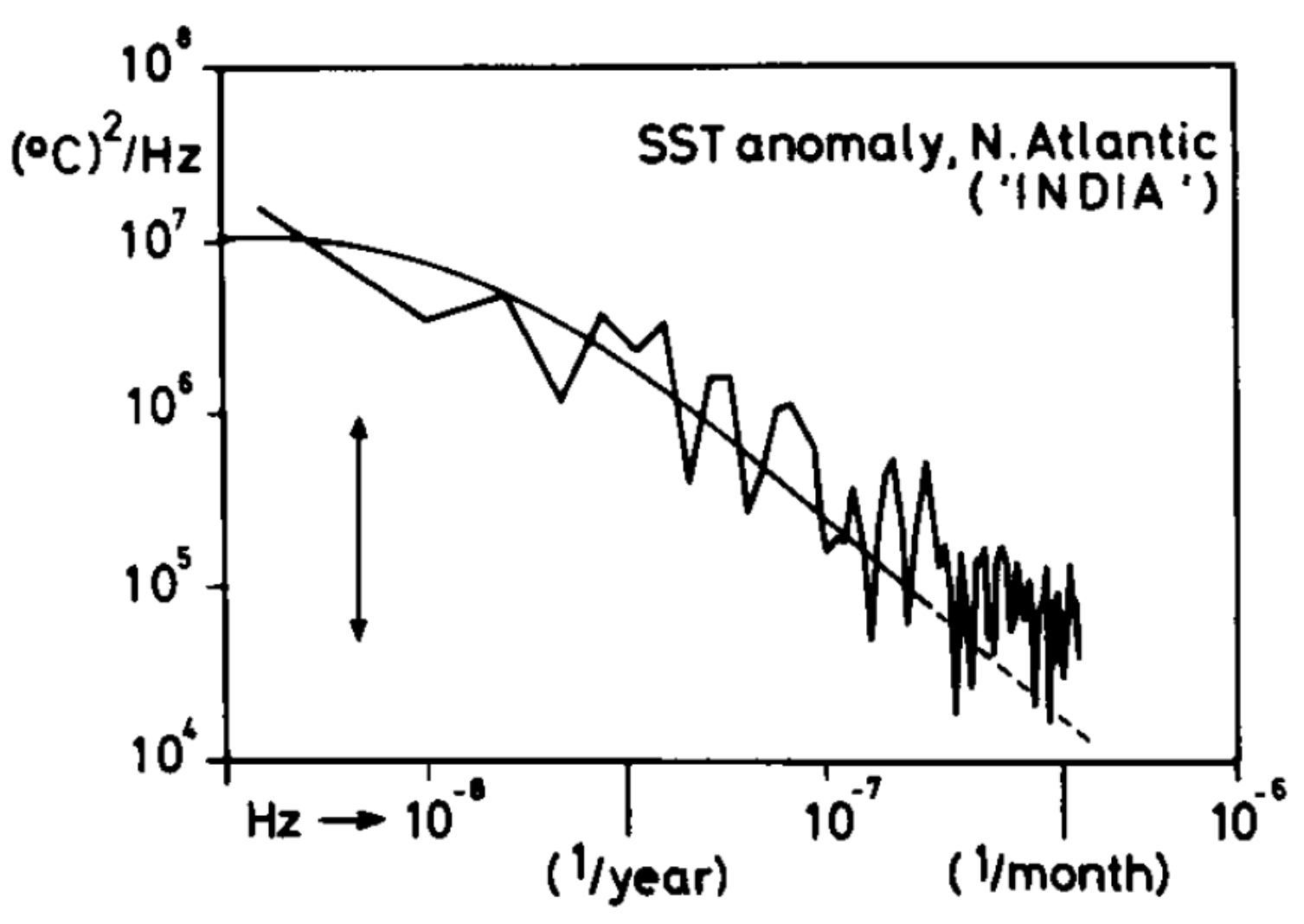}
    \caption{\small Spectrum of Sea Surface Temperature (SST) anomaly at Atlantic Ocean Weather Ship India for the period 1949-1964. The arrows indicate 95\% confidence interval. The smooth curve was calculated from Hasselmann's stochastic model of SST variability which successfully explains the ubiquitous red-noise behaviour of the SST signal. From \citeauthor{Hasselmann1977}\cite{Hasselmann1977}, licensed under a Creative Commons Attribution (CC BY) license.}
    \label{fig:Hassel_RedNoise}
\end{figure}

\begin{equation}
    \frac{\partial p}{\partial t}+\frac{\partial}{\partial y_i}(\hat{v_i}p)-\frac{\partial}{\partial y_i}\left(D_{ij}\frac{\partial p}{\partial y_j}\right)=0,
    \label{eq:4}
\end{equation}
where 
\begin{equation}
    \hat{v_i}=\left<v_i\right>-\frac{\partial}{\partial y_j} D_{ij} = \left<v_i\right>-\pi\frac{\partial}{\partial y_j}F_{ij}(0)
\end{equation}
Therefore, the model includes indirect feedback through the dependence of the diffusion coefficients on the climatic state. This stabilising feedback mechanism is vital for formulating a realistic model without which climatic fluctuations will grow unboundedly. The presence of the diffusion term leads to a spread in the probability distribution of the climate state $p$ at a later time even for a well-defined initial state, thus implying that the climate evolution is rather a stochastic process than a deterministic one.  However, the internal feedback terms imply that the climate system does have a finite degree of predictability despite its stochasticity.  


Hasselmann's stochastic model was successful in explaining the ubiquitous red noise behaviour observed in the long-term Sea Surface Temperature (SST) anomaly data (Fig.~\ref{fig:Hassel_RedNoise}) without incorporating external perturbations\cite{Hasselmann1977}. In this case, the evolution of the SST anomaly $y(t)$ is governed by Eq. \ref{eq:2}, where the forcing function $v({\bf x}, {\bf y})$ is determined by the random fluxes of heat and momentum across the air-sea interface. Thus, the red noise spectrum of the SST anomaly can be interpreted naturally as the response of the oceanic surface layers to short-time-scale atmospheric forcing which acts as a white-noise generator.

\subsection{ Hasselmann and Manabe's Contributions to the Development of Coupled General Circulation Models}

Hasselmann realized the importance of coupling the global ocean and atmospheric GCMs together, along with other climate sub-systems which interact with them, in order to build a more realistic climate model that can be used for more accurate forecasting. The ocean-GCMs were strongly sensitive to minor changes in the surface forcing. Such fluctuations could potentially lead to large climate variability down the line, as implied by the stochastic model. Hence, it was essential to model the dynamical processes at the air-sea interface as accurately as possible in the global coupled ocean--atmospheric GCMs (CGCMs) that were under development. 

In 1988, the WAMDI group, comprising Hasselmann and others, constructed a third-generation surface wave model, wherein the evolution of the wave spectrum was inferred from the basic equations governing the spectral energy balance rather than ad hoc assumptions about its shape\cite{WAM1988}. Its output agreed well with observational data from various sources under different situations. 

Since the oceans act as carbon sinks, changes in the ocean circulation can significantly affect the $\rm{CO_2}$ concentration in the atmosphere, as demonstrated by the 3D carbon cycle model of Heinze \textit{et al.}\cite{Heinze1991}, based on a 3D ocean transport model\cite{Hasselmann1987}. Thus, the climate dynamics 
are described by a coupled ocean--atmosphere--surface--wave--carbon-cycle model. 

Such a model was developed in 1980, when \citet{Manabe1980} coupled a GCM of the atmosphere with a heat and water balance model of the continents, and a mixed layer model of the oceans. 
\citet{Manabe1980} studied the model's response to the quadrupling of the CO$_2$ concentration in the atmosphere. Their model succeeded in reproducing the large-scale characteristics of seasonal and geographical variation of the observed atmospheric temperature. 

In 1992, Cubasch, Hasselmann \textit{et al.}\cite{Cubasch1992} simulated the climate changes caused by anthropogenic emission of greenhouse gases during the next 100 years, for the IPCC Scenarios A ("business as usual") and D ("accelerated policies"), using one of the Hamburg CGCMs. The model simulated the response of the oceans to stochastic forcing by coupling the numerical weather forecast model of the atmosphere with the ocean model appropriate for describing large-scale geostrophic circulation via the air-sea fluxes of momentum, sensible and latent heat, short and long wave radiation and fresh water. Three time-dependent greenhouse warming simulations (IPCC Scenarios A and D and, a $\rm{CO_2}$ doubling experiment) were carried out with the CGCM, along with a control run to distinguish anthropogenic warming from the natural variability of the model. 
In their model, the near-surface temperature increased by 2.6 K in Scenario A and by 0.6 K in Scenario D in 100 years. However, during the first 10-50 years, their estimates were lower than the corresponding IPCC estimates, which were computed from the relatively simpler box-diffusion-upwelling model (see Fig. 10 in \citet{Cubasch1992}). This warming delay could be explained by the "cold start" to the model (lack of a warm-up period before the start of simulations). It could also be due to a stronger heat uptake by the oceans, as observed in the deep oceans in higher latitudes. This might have resulted from a detailed description of the deep ocean in the model. At the later stages of all three warming simulations, it is possible to easily distinguish the global patterns of climate change in the former from the observed natural variability of the control run. In the $\rm{CO_2}$-doubling run, one is clearly able to distinguish between two time scales -- an initial rapid increase over an $\approx 5$-year period followed by a more gradual increase over the remaining duration. This is consistent with the circulation time scales of the upper and abyssal ocean layers, respectively. The sea-level rise predicted from this model is also smaller than the corresponding IPCC estimates. This is suggestive of a delay in the sea-level response to thermal expansion and, was consistent with previous estimates. Later, Joos \textit{et al.}\cite{Joos2001} investigated global warming feedback on terrestrial carbon uptake under the IPCC emission scenarios by coupling a dynamic global vegetation model with the CGCMs, formulating a coupled physical-biogeochemical climate model.


\section{\label{sec:Stochastic}{{Parisi's Contributions to Understanding Noise-induced Effects in our Climate}}}

The analysis of paleoclimatic records over the last million years has shown the occurrence of major climate changes with an apparent periodicity of $10^5$ years and temperature variations\cite{Mason1976} of the order of 10 K. The transitions occur as an alternation between two stable climatic equilibrium states which are often interpreted as glacial and interglacial climate\cite{Hays1976}. The two state equilibria of the climate system can modeled by studying the effect of changes in the annually averaged solar radiation on the global earth temperature $T$ using a zero-dimensional Budyko-Sellers model\cite{Budyko1969,Sellers1969,North1975},  
\begin{equation}
\label{Albedo}
    C\frac{dT}{dt}=R_{in}(T)-R_{out}(T) \implies \frac{dT}{dt}=F(T)
\end{equation}
This belongs to the class of deterministic energy balance models which are the simplest possible models of the climate system. Here, $C$ is the thermal capacity of the earth, $R_{in}$ is the incoming solar radiation and $F(T)=(R_{in}-R_{out})/C$. Based on our present knowledge of the ice albedo feedback as a function of the temperature, the outgoing solar radiation $R_{out}$ can be parameterized as the sum of the reflected part $\alpha(T)R_{in}$, where $\alpha(T)$ is the globally averaged albedo, and the emitted infrared radiation $\epsilon(T)$. The fixed points of Eq. \ref{Albedo} represent the steady states or the 'climates' of the model. The effective potential $V(T)=-\int F(T)dT$ of the system in Eq. \ref{Albedo} is postulated to be bistable\cite{Ghil1976,BhattacharyaGhil1979,North1979}, i.e., having one unstable ($T_2$) and two stable steady states ($T_1$ and $T_3$), based on the two ``observed climates'' from the paleoclimatic records.

\begin{figure}
    \centering
    \includegraphics[scale=0.35]{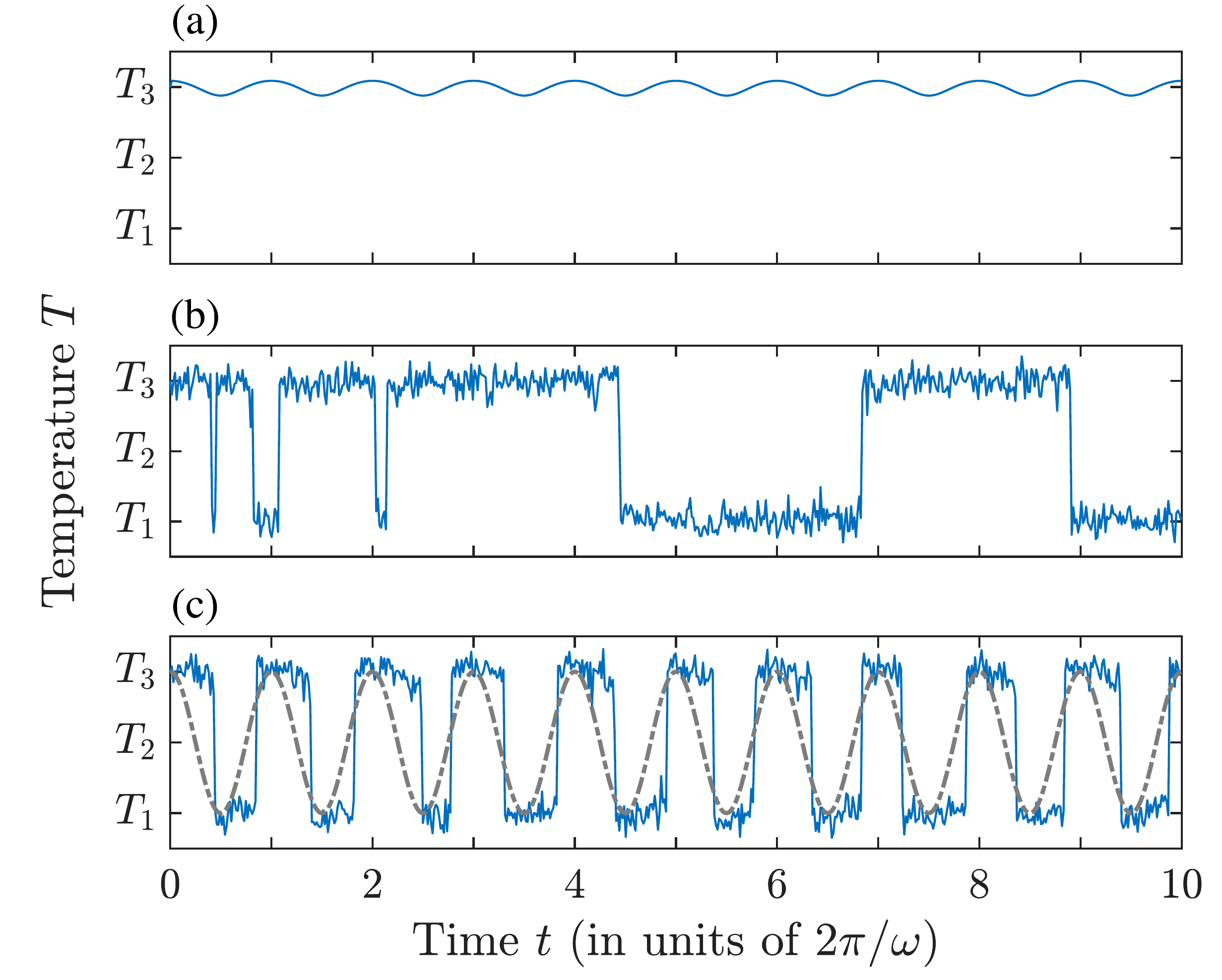}
    \caption{\small Schematic diagram of the response of global mean temperature to the effect of (a) external periodic forcing (Milankovitch forcing) with angular frequency $\omega$ in the absence of noise, (b) internal noise (arising due to atmospheric and oceanic circulations) in the absence of external periodic forcing, and (c) a combination of both. The temperatures $T_1$ and $T_3$ correspond to the stable steady states of the system, while $T_2$ represents the unstable one. It is only in case (c) that we observe noise-induced periodic transitions between the two stable climatic states with the angular frequency $\omega$. The figure was obtained from simulations of the stochastic dynamical system proposed in 1981 by Benzi \textit{et al.}\cite{Benzi1981}, which exhibits stochastic resonance.}
    \label{fig:benzi_parisi}
\end{figure}

\begin{figure}
    \centering
    \includegraphics[scale=0.55]{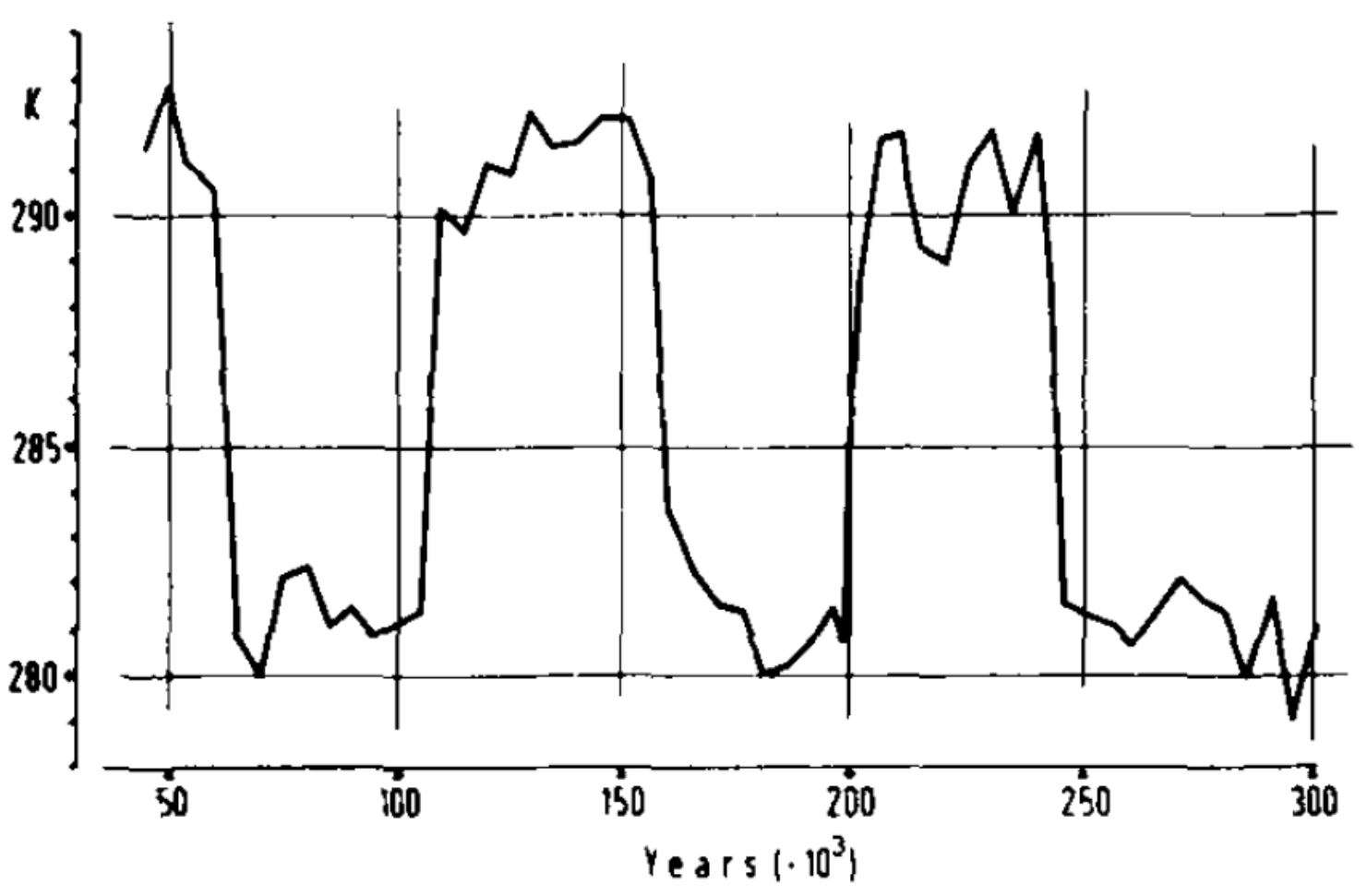}
    \caption{\small Simulated temperature response of the stochastically perturbed climate model including the effects due to Milankovitch forcing. Periodic transitions were seen every $10^5$ years between two stable climate regimes differing by 10 K in their mean temperatures. From \citeauthor{Benzi1982}\cite{Benzi1982}, licensed under a Creative Commons Attribution (CC BY) license.}
    \label{fig:benzi_stochastic}
\end{figure}

In 1930 Serbian scientist Milutin Milankovitch\cite{Milankovitch1930} related these transitions to changes in global solar radiation caused by periodical changes in the Earth’s orbital parameters, thereby associating climate change with external astronomical forcing. The effect on the solar radiation due to the periodic variation of Earth's orbit eccentricity can be incorporated in Eq. \ref{Albedo} by parameterizing the incoming solar radiation as $R_{in}=Q(1+A\textrm{cos}\omega t)$, where $Q$ is the solar constant, $A=5\times 10^{-4}$ is the amplitude of the Milankovitch forcing and $\omega=2\pi/10^5$ yr$^{-1}$ is the angular frequency of this periodic variation.  However, it was found through actual calculations\cite{schneider_thompson_1979} that at most 1 K variation in temperature is caused by the Milankovitch forcing, which is much less than what is observed from climate records (see Fig.~\ref{fig:benzi_parisi}a). 
Using Klaus Hasselmann's concept\cite{Hasselmann1976} of modelling short-time scale phenomena as stochastic perturbations, in 1981 Alfonso Sutera\cite{Sutera1981} quantitatively showed that finite amplitude stochastic disturbances in an energy balance model could lead to transitions between the equilibria of the model. However, the transitions between climate states occur randomly and not periodically, with a characteristic time of $10^5$ years (see Fig.~\ref{fig:benzi_parisi}b for a schematic illustration). In 1982 Roberto Benzi, Giorgio Parisi and coworkers\cite{Benzi1982,Benzi1983} hinted that the problem may be solved by incorporating the effects of internal noise arising due to atmospheric and oceanic circulations along with the periodic changes in the solar radiation due to Milankovitch forcing into the climate model (Fig.~\ref{fig:benzi_parisi}c). The reason being that the combined effects of  internal noise and external periodic forcing can result in almost periodic behaviour. Eq. \ref{Albedo} can then be modified as 
\begin{equation}
    dT=F(T)dt+\sigma dW
\end{equation}
where $\sigma^2$ is the variance of the stochastic perturbation, and $dW$ are the infinitesimal increments of a Wiener process which is the mathematical analogue of the standard Brownian motion. By both numerically simulating the above climate model\cite{Benzi1982} and using the expression derived by Kramers\cite{Kramers} for the mean time of transition\cite{Benzi1983} between the two stable states of a double-well potential, Benzi, Parisi and collaborators were able to reproduce the nearly periodic transitions between two climate states, differing in the global mean temperature by 10 K (i.e., $T_3-T_1\approx 10$ K), occurring at intervals of $10^5$ years (Fig. \ref{fig:benzi_stochastic}). 
 The response of the model appeared to be considerably enhanced when the characteristic time of the small-scale fluctuations matched closely with the period of the external force, similar to what happens during a typical resonance (when the frequency of the external force matches with the natural frequency of oscillations) and they called it "stochastic resonance"\cite{Benzi1981,Benzi1982,Benzi1983}. Similar results were obtained by Nicolis\cite{Nicolis}, both analytically and numerically, from the Fokker-Planck equation derived from the stochastic climate model.  

Such abrupt transitions between coexisting states or attractors have been referred to as \textit{tipping points}\cite{Lenton2008,NoiseTipping}. It is fascinating that the Earth's climate can exhibit abrupt \textit{noise-induced tipping} from one climate regime to another under the influence of stochastic fluctuations due to internal mechanisms. 


\section{Climate change due to Anthropogenic factors}

As seen in the previous section, the combined effects of the internal variability of the Earth's climate and an external astronomical periodic forcing may be responsible for nearly regular tipping between glacial and inter-glacial climate states. However, since the climate is governed by a myriad of intricate interactions within the coupled ocean-atmosphere-cryosphere-land system, perturbations to any of these due to anthropogenic factors (such as the increase in greenhouse gas concentration due to the burning of fossil fuels) could amplify with time and the Earth system could cross a tipping point, which could cause large-scale impacts on human and ecological systems\cite{Lenton2008}. Both Manabe and Hasselmann made important contributions to the study of the consequences of a changing climate on the tipping elements (e.g. the Arctic sea ice, Atlantic thermohaline circulation, boreal forest, etc.),  and to identify the role of human activities on them.

\subsection{Consequences of enhanced concentrations of greenhouse gases in the atmosphere}

\begin{figure}
    \centering
    \includegraphics[scale=0.7]{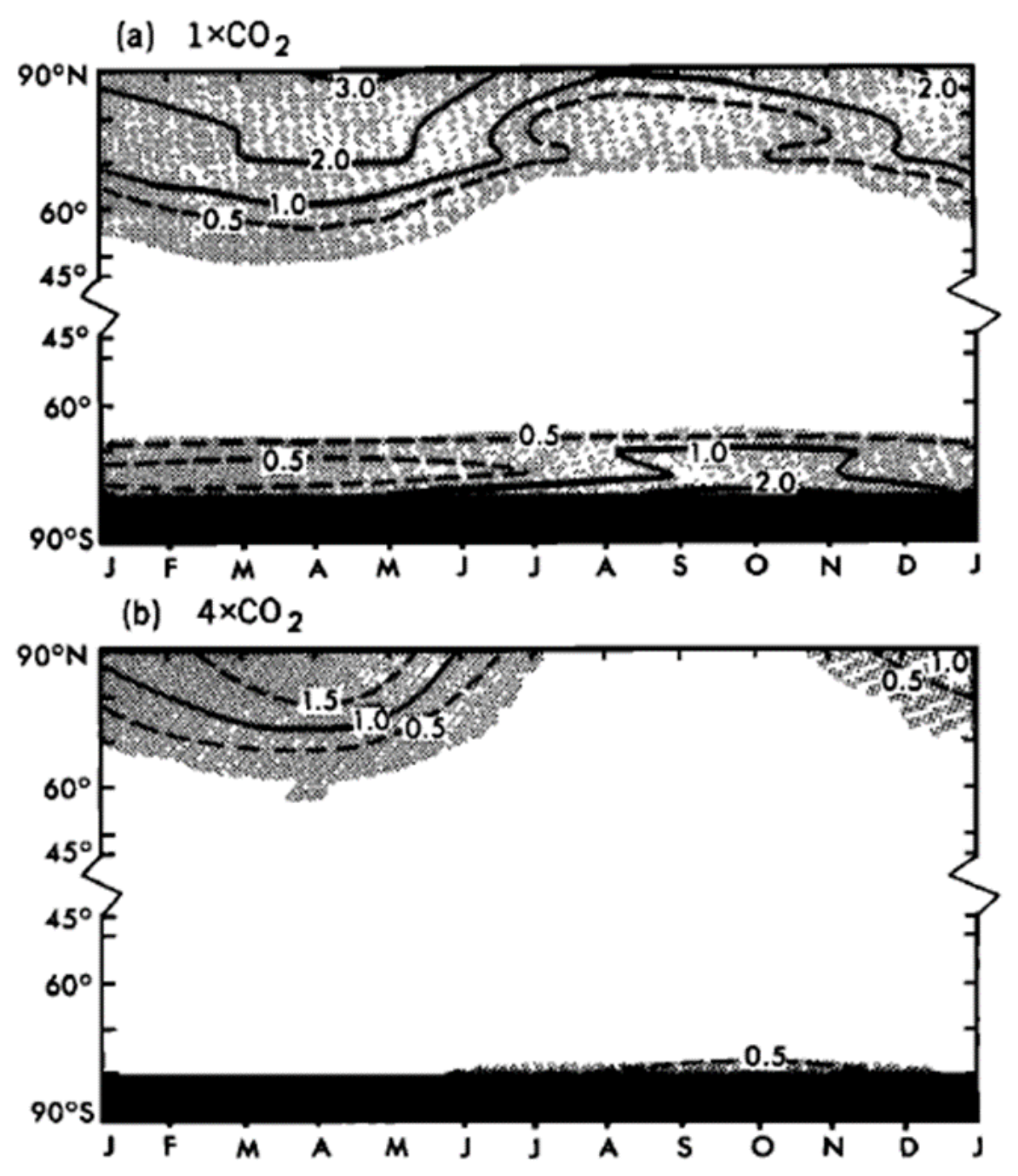}
    \caption{\small Temporal variation in the distribution of sea ice thickness (in centimetres) across different latitudes for 1$\times$CO$_2$ (\textit{top}) and 4$\times$CO$_2$ (\textit{bottom}) experiments. Regions where sea ice thickness exceed 0.1 m are shaded.  From \citeauthor{Manabe1980}\cite{Manabe1980}. Published in 1980 by the American Geophysical Union.}
    \label{fig:seaice}
\end{figure}

 \citet{Manabe1980}, in 1980, were able to show through their coupled ocean-atmosphere model that around the Arctic Ocean the warming of the atmosphere surface layer due to increased CO$_2$ concentrations would be much larger in the winter than in the summer (Fig. \ref{fig:seaice}). This finding was attributed to the reduction of sea ice thickness. This reduction subsequently reduces the albedo, which then leads to a substantial increase of the net incoming radiation in summer. However, this surplus of energy does not bring a remarkable increase of the average summer temperature, because it is absorbed by the oceans or used for melting the sea ice. Thus, during the coming winter season the sea ice is delayed and/or reduced in thickness, subsequently reducing its thermal insulation effect. As during winter the air-sea temperature differences become large, this finally leads to an increased warming of the surface atmospheric layer. The winter warming is enhanced further by the stable stratification of the model atmosphere in the winter, which confines the warming to the surface layers.

\begin{figure}
    \centering
    \includegraphics[scale=0.5]{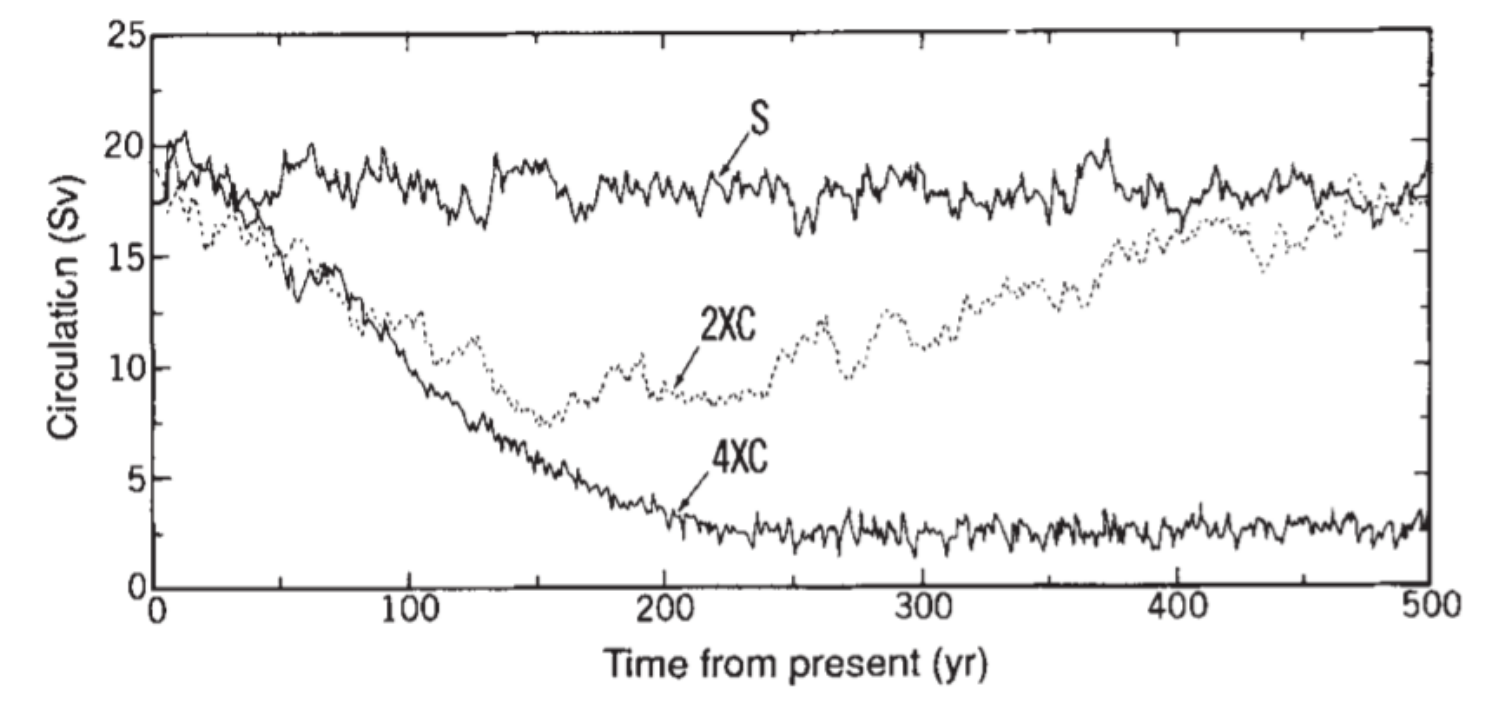}
    \caption{\small Temporal variation of the intensity of the thermohaline circulation in the North Atlantic from the standard integration (S) in which the level of CO$_2$ is unchanged, on doubling the concentration of atmospheric CO$_2$ (2XC) and then quadrupling the concentration of atmospheric CO$_2$ (4XC) and the. Here the intensity is defined as the maximum value of the stream function representing the meridional circulation in the North Atlantic. From \citeauthor{Manabe1994}\cite{Manabe1994}. Published 1994 by the American Meteorological Society.}
    \label{fig:thermohaline}
\end{figure}

Later, in \citeyear{Stouffer1989},
Manabe and his colleagues demonstrated an interhemispheric asymmetry in temperature changes under an increase of CO$_2$ and described how it is affected by oceanic processes\cite{Stouffer1989}: the warming of surface air was predicted to be faster in the Northern Hemisphere than in the Southern Hemisphere. At the same time, the warming over the northern North Atlantic was predicted to be relatively slow because of the weakening of the large-scale ocean circulation (thermohaline circulation), which acts as a countereffect. In fact, a few years later Manabe and Stouffer showed the non-linear effects of increasing CO$_2$ concentrations (unchanged vs doubling vs quadrupling) to the thermohaline circulation (Fig. \ref{fig:thermohaline})\cite{Manabe1993,Manabe1994}. The thermal and dynamical structure of the oceans changes markedly in the quadrupled-CO$_2$ climate: the ocean settles into a new stable state in which the thermohaline circulation has ceased entirely, and the thermohaline deepens substantially. These changes prevent the ventilation of the deep ocean and could have a profound impact on the carbon cycle and biogeochemistry of the coupled system.

In \citeyear{Joos2001}, Hasselmann and coworkers used a coupled physical-biogeochemical model to investigate the climate-land biosphere feedbacks under a range of projected greenhouse gases emission scenarios\cite{Joos2001}, and demonstrated that the terrestrial system can generate a substantial response to climatic shifts within a few decades. Their model suggests that an enhanced warming could lead to a large-scale dieback of the boreal forests, with a continuous transition to temperate deciduous forests or open grasslands without abrupt tipping point behaviour. On the other hand, the warming in high northern latitudes leads to a gradual transition from tundra to boreal forests in North America and Siberia. 

Hasselmann and coworkers\cite{Hoos2001} also developed a nonlinear impulse response function of the coupled carbon cycle-climate system and computed the temporal evolution of the spatial patterns in selected impact-relevant climate fields under long-term CO$_2$ emission scenarios. They predicted that if all the estimated fossil fuel resources are burnt, the climate system will be carried into a range of extreme CO$_2$ concentrations leading to increase in temperatures and rise in sea level, whose magnitudes are far beyond the calibration ranges of existing climate models. Even a freezing of emission levels would limit global warming in the long run. Two years later, Hasselmann and collaborators\cite{Hasselmann2003} used an integrated assessment model consisting of the nonlinear impulse response climate model coupled to an elementary economic model\cite{Hasselmann1997_1} 
and showed that, owing to the long term memory of the climate system, major climate change could be avoided only by reducing global emissions to a small fraction of present levels within one or two centuries.

\subsection{Hasselmann's contribution to the identification of the climate change signal in observed data}

Although there was growing qualitative and circumstantial evidence of global warming due to increasing atmospheric greenhouse gas concentrations predicted by the then state-of-the-art CGCMs, there was a lack of a quantitative measure to distinguish climate change signal from background noise of natural climate variability in observed climate data. 

In 1979 Hasselmann identified the problem as a pattern-detection problem\cite{Hasselmann1979}. He demonstrated that the unfiltered atmospheric response to an external forcing mechanism, inferred from observational or model data, would fail any significance test for pattern-detection. Hence he proposed a way of constructing a set anticipated response patterns or "guess patterns" which could be used as a basis for re-constructing the observed response pattern, so as to increase the signal-to-noise ratio. Later, he derived an optimal linear filter or "fingerprint", in order to detect time-dependent multivariate climate change signal in the presence of natural climate variability\cite{Hasselmann1993}.

\begin{figure}
    \centering
    \includegraphics[scale=0.6]{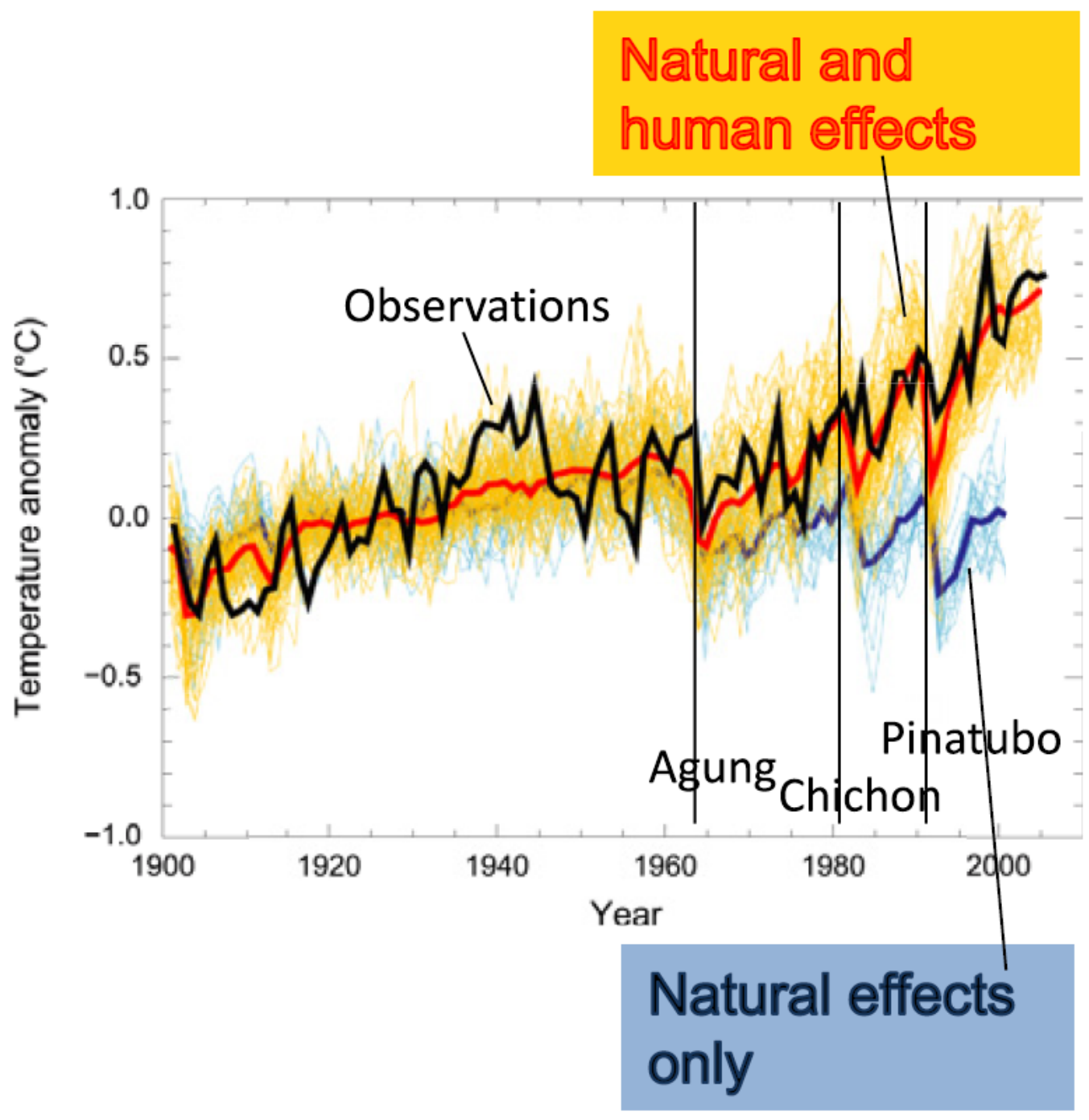}
    \caption{\small Global average warming from observational data (black), and from an ensemble of climate model simulations that include only natural forcing (blue), and both human and natural forcing (orange). Individual model simulations are shown by thin lines, while thick lines indicate their average. The effects of strong volcanic eruptions are marked by vertical bars.  From Hegerl \textit{et al.}\cite{Hegerl_2011}. \copyright\,2022 IOP Publishing. Reproduced with permission of IOP Publishing Limited through PLSclear.}
    \label{fig:hegerl}
\end{figure}

Through the method of \textit{optimal fingerprinting}, Hasselmann and his colleagues first established that climate change has occurred over the last 30 years, the cause of which cannot be singly ascribed to natural climate variability\cite{Hegerl1996}. In subsequent works, they additionally demonstrated that the detected climate change signal can be most likely attributed to an increase in greenhouse gas concentrations, rather than some other external forcing mechanisms, such as changes in the solar constant, volcanic activity or modified land-use practices\cite{Hasselmann1997,Hegerl_2011}(Fig.~\ref{fig:hegerl}).


\section{A complex future}

The recognition of the contributions of Syukuro Manabe, Klaus Hasselmann and Giorgio Parisi to climate science and complex systems by the Nobel committee comes at a critical juncture when the world is on the verge of a climate crisis. The pioneering works of the Physics Nobel Laureates of 2021 provide a deep insight into the past, present and future of our climate and the associated role of human activities. 

Syukuro Manabe and Klaus Hasselmann established the foundation on which models for weather prediction stand today. The very recent “Destination Earth” initiative\cite{Twin2021}, which aims to digitally replicate the state and temporal evolution of the Earth system with available observations and the laws of physics, in order to monitor and predict extreme weather events and climate change, is based on such models. 

Additionally, stochastic parameterization\cite{StochasticParameterization} of weather models, i.e., parameterization that adapts automatically to different spatial scales, is an active direction of research for seamless predictions of weather and climate in the foreseeable future. Many such stochastic parameterizations are based on Hasselmann's\cite{Hasselmann1976} ingenious idea of separating physical processes by time scales. As there is often a direct relationship between spatial and temporal scales of variability in geophysical systems, such temporal scale separations can then help in decomposing small-scale features from large-scale phenomena.

The large natural variability of the climate system on different timescales is highly susceptible to relatively small changes in the natural or anthropogenic forcing. Despite the advancements in numerical weather and climate prediction models, the multiscale nature of the climate system implies that many important physical processes have not yet been resolved. Importantly, more research is needed to improve our understanding of the interactions between different tipping elements. Recent developments in dynamical systems theory and non-equilibrium statistical physics, in addition to innovative data analysis methods offer promising approaches\cite{Ghil_Review,networkTipping,Volcanicforcing,physrep2021}.

In particular, the complex network approach~\cite{epjst2009,networksinclimate} assumes that climate phenomena can be understood, at least in part, by using networks or graphs, where the nodes represent geographical regions and the links represent causal interactions between phenomena occurring in different regions. The representation of the Earth system in terms of interacting networks, or networks-of-networks, shields light on how changes in one network that represents a sub-system of the Earth system affects other sub-systems. Climate network studies can also contribute to project future changes, to provide uncertainty estimates, and to identify warning signals of the closeness to tipping points~\cite{grl2013}. The network approach is particularly attractive due to its inherent nonlinear and data-integrative nature and can arguably complement numerical modelling methods~\cite{Ludeschere1922872118}.

These new advances may shed light on the presence of critical tipping points, how close we are to them, and how crossing them will affect the future of the Earth system, for example, the changing of the Amazon forest into savanna~\cite{Amazon1,Amazon2} or the slowing of the thermohaline circulation in the Atlantic oceans~\cite{Atlantic1,Atlantic2}. Moreover, these new models will allow to better simulate the effects of possible mitigation actions that can be implemented, as for example geoengineering for capturing CO$_2$. There is also a growing need to not only assess the economic impact of climate change but also to find solutions for the economy that can help to transfer economic losses associated with climate change into opportunities for creating a climate-friendly sustainable economy. This will aid political decisions by providing a better estimation of the uncertainities involved. Including nonlinearities and stochasticities in coupled climate-economy models\cite{Economic1,Economic2}, and using financial macro-networks\cite{STOLBOVA2018239} to comprehensively evaluate the economic impact of climate policies are promising lines of research in this direction.

Global challenges demand global solutions. Only together we can build a better and more sustainable planet for all its present and future inhabitants. This year's Nobel Prize is a great reminder that advancing our knowledge about our planet requires large-scale concerted efforts -- improved research across various disciplines of the scientific community to provide a scientific basis for decision making, and joint actions by governments, companies and citizens -- for mitigating our negative impacts on our planet~\cite{Hasselmann2003}.

\begin{acknowledgments}
All authors received funding from the European Union’s Horizon 2020 research and innovation programme under the Marie Sklodowska Curie Grant Agreement No 813844. C.M. also acknowledges funding from the Spanish Ministerio de Ciencia, Innovación y Universidades (PGC2018-099443-B-I00 ) and the ICREA ACADEMIA program of Generalitat de Catalunya. J.K. was supported by the Russian Ministry of Science and Education Agreement No. 075-15-2020-808.
\end{acknowledgments}

\section*{Data Availability Statement}
No data has been generated in this study.

\nocite{*}
\bibliography{aipsamp}

\end{document}